\documentclass[twocolumn, trackchanges]{aastex631}
\usepackage{amsmath}
\usepackage{subfigure}
\usepackage{wasysym}
\usepackage{comment}
\usepackage{multirow} 
\usepackage{booktabs}
\newcommand{\teff}{{\rm T_{eff}}}

\shortauthors{Louden et. al}
\shorttitle{Hot Stars Are Misaligned}

\begin{document}

\title{A Larger Sample Confirms Small Planets Around Hot Stars Are Misaligned 
\footnote{The data presented herein were obtained at the W. M. Keck Observatory, operated as a scientific partnership among the California Institute of Technology, the University of California, and the National Aeronautics and Space Administration. The Observatory was made possible by the generous financial support of the W.M.~Keck Foundation.}}

\correspondingauthor{Emma M. Louden}
\email{emma.louden@yale.edu}

\author[0000-0003-3179-5320]{Emma M. Louden}
\affiliation{Department of Astronomy, Yale University, 52 Hillhouse Avenue, New Haven, CT 06511, USA}

\author[0000-0002-7846-6981]{Songhu Wang} 
\affiliation{Department of Astronomy, Indiana University, 727 East 3rd Street, Bloomington, IN 47405-7105, USA}

\author[0000-0002-4265-047X]{Joshua N.\ Winn} 
\affiliation{Department of Astrophysical Sciences, Princeton University, 4 Ivy Lane, Princeton, NJ 08544, USA}

\author[0000-0003-0967-2893]{Erik A.\ Petigura} 
\affiliation{Department of Physics \& Astronomy, University of California Los Angeles, Los Angeles, CA 90095, USA}

\author[0000-0002-0531-1073]{Howard Isaacson} 
\affiliation{501 Campbell Hall, University of California at Berkeley, Berkeley, CA 94720, USA}\affiliation{Centre for Astrophysics, University of Southern Queensland, Toowoomba, QLD, Australia}

\author[0000-0002-9305-5101]{Luke Handley} 
\affiliation{Department of Physics \& Astronomy, University of California Los Angeles, Los Angeles, CA 90095, USA}
\affiliation{Department of Astronomy, California Institute of Technology, Pasadena, CA 91125, USA}

\author[0000-0001-7961-3907]{Samuel W.\ Yee} 
\affiliation{Department of Astrophysical Sciences, Princeton University, 4 Ivy Lane, Princeton, NJ 08544, USA}
\affiliation{Center for Astrophysics \textbar \ Harvard \& Smithsonian, 60 Garden St, Cambridge, MA 02138, USA}

\author[0000-0001-7708-2364]{Corey Beard} 
\altaffiliation{NASA FINESST Fellow}
\affiliation{Department of Physics \& Astronomy, The University of California, Irvine, Irvine, CA 92697, USA}

\author[0000-0001-8898-8284]{Joseph M. Akana Murphy} 
\altaffiliation{NSF Graduate Research Fellow}
\affiliation{Department of Astronomy and Astrophysics, University of California, Santa Cruz, CA 95064, USA}


\author[0000-0002-3253-2621]{Gregory Laughlin} 
\affiliation{Department of Astronomy, Yale University, 52 Hillhouse Avenue, New Haven, CT 06511, USA}


\begin{abstract}
The distribution of stellar obliquities provides critical insight into the formation and evolution pathways of exoplanets.  In the past decade, it was found that hot stars hosting hot Jupiters are more likely to have high obliquities than cool stars, but it is not clear whether this trend exists only for hot Jupiters or holds for other types of planets. In this work, we extend the study of the obliquities of hot (6250-7000\,K) stars with transiting super-Earth and sub-Neptune-sized planets. We constrain the obliquity distribution based on measurements of the stars' projected rotation velocities. Our sample consists of 170 TESS and \textit{Kepler} planet-hosting stars and 180 control stars chosen to have indistinguishable spectroscopic characteristics. In our analysis, we find evidence suggesting that the planet hosts have a systematically higher $\langle \sin i \rangle$ compared to the control sample. This result implies that the planet hosts tend to have lower obliquities. However, the observed difference in $\langle \sin i \rangle$ is not significant enough to confirm spin-orbit alignment, as it is 3.8$\sigma$ away from perfect alignment. We also find evidence that within the planet-hosting stars there is a trend of higher obliquity (lower $\langle \sin i\rangle$) for the hotter stars ($\teff > 6250$ K) than for the cooler stars in the sample. This suggests that hot stars hosting smaller planets exhibit a broader obliquity distribution($\langle \sin i\rangle = 0.79 \pm 0.053$) than cooler planet-hosting stars, indicating that high obliquities are not exclusive to hot Jupiters and instead are more broadly tied to hot stars.
\end{abstract}


\section{Introduction} \label{sec:intro}

The stellar obliquity, defined as the angle between the spin vector of a star and the orbital angular momentum vector of its planets, is one of the fundamental parameters used to describe the geometry of an exoplanetary system. Stellar obliquity is of particular consequence to studies of the formation and evolution mechanisms of planetary systems, as stellar obliquity drives dynamical interactions and constrains the formation pathways for exoplanets. 

The Sun's obliquity has been determined through helioseismology to be $7^\circ$ \citep{2005ApJ...621L.153B}, indicating reasonable alignment between the plane of the Solar System's orbital angular momentum and the Sun's spin axis. However, measurements of stellar obliquities in exoplanetary systems suggest that such alignment may not always be the norm \citep{albrecht_stellar_2022, 2015ARA&A..53..409W, triaud_rossitermclaughlin_2017}.

Recent observations of systems such as KELT-11 \citep{cegla_exploring_2023}, HAT-P-49 \citep{bourrier_dream_2023}, TOI-1859 \citep{dong_toi-1859b_2023}, and TOI-1842 \citep{hixenbaugh_spin-orbit_2023} add to a growing body of planets in misaligned orbits around their stars. These obliquities were constrained with the Rossiter-McLaughlin (RM) \citep{mclaughlin_results_1924, rossiter_detection_1924} method and rely upon the ability to observe the radial velocity of the star while the planet is transiting. 

The frequent occurrence of hot Jupiters in misaligned orbits around hot stars, as observed by \citet{schlaufman_evidence_2010} and \citet{winn_hot_2010}, has led to a critical examination of the intrinsic relationship between these misalignments and the characteristics of the host stars and their planets. In contrast to these observed misalignments (higher obliquities), current RM measurements for warm Jupiters – defined as giant planets with $a/R_* > 12$ \citep{wang_aligned_2021} – are mostly on cooler stars and typically find lower obliquities \citep{rice_tendency_2022, harre_orbit_2023, rice_soles_2021, wang_aligned_2021, wright_soles_2023}. Spin-orbit alignment (low obliquity) is also observed in multiplanet systems orbiting cooler stars \citep{albrecht_low_2013, wang_stellar_2018, zhou_warm_2018, dai_toi-1136_2023,wang_aligned_2022,rice_evidence_2023, lubin_toi-1670_2023}. The relative small number of RM measurements on systems with hot stars and planets other than hot Jupiters prevents us from determining whether the misalignment (higher obliquity) is a unique characteristic of hot Jupiters orbiting hot stars or if it is a broader phenomenon applicable to planets of various sizes orbiting hot stars. Resolving this ambiguity is vital for advancing our understanding of the formation and evolutionary trajectories of exoplanetary systems. 

The resolution would be straightforward if we could expand the current RM measurements to different types of planets around hot stars. However, most current measurements are obtained for short-period gas giants around F, G, and K stars for practical reasons: they are easy to measure via the RM method because of their large radius and proximity to their host star. Additionally, we know of fewer small transiting planets around hot stars because the transit SNR is lower and because the hot stars are often rapidly rotating, which inhibits Doppler confirmation. While the advent of extreme precision spectroscopy permits RM measurements of individual systems with smaller planets (e.g., \cite{2023NatAs...7..198Z}), the progress in assembling a sample of RM measurements large enough for statistical analysis is slow.

\begin{figure}[ht!]
    \centering    \includegraphics[width=0.45\textwidth]{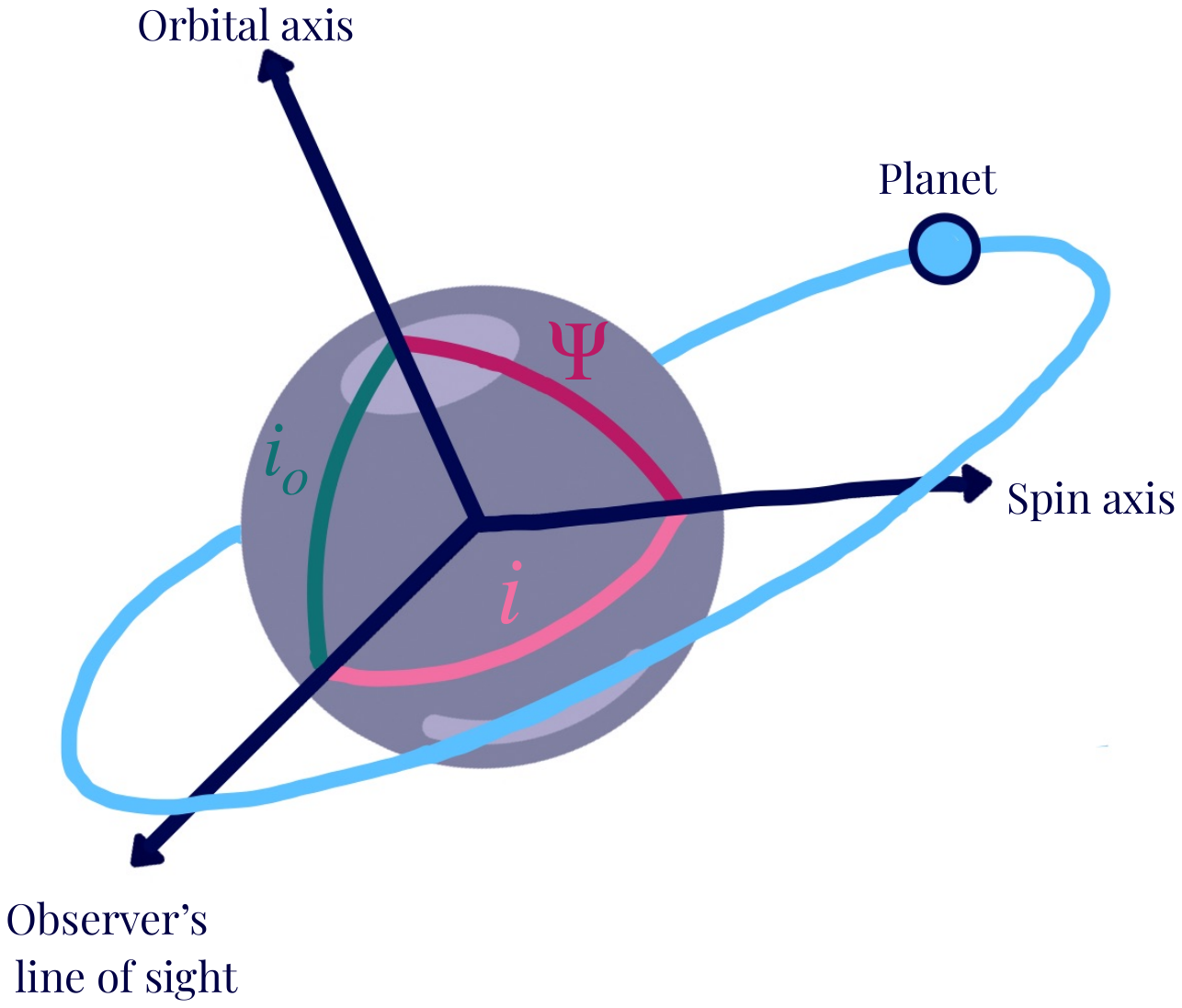}
    \caption{\textbf{Planetary Geometry.} The angle $\Psi$ is the full stellar obliquity. The angle $i_0$ is the inclination between the planet's orbital axis and the observer's line of sight. This is near $90\deg$ for transiting planets. $i$ is the star's inclination.}
    \label{fig:Obliquity}
\end{figure}

An alternative to the RM method is the use of the $v\sin{i}$ technique to constrain the obliquity along the line of sight. This method was first employed by \cite{2010ApJ...719..602S}. In the context of transiting planets, one degree of freedom, the orbital inclination $i_0$, is constrained by the requirement that for transiting planets $i_0 \approx 90$ (see Figure \ref{fig:Obliquity}). Thus, by comparing the $v\sin{i}$ distributions of transiting planet hosts with those of a control sample — which ideally has the same distribution of rotation velocities as the planet hosts — we can gain valuable insights into the distribution of obliquities. 

\cite{2017AJ....154..270W} employed the $v\sin{i}$ technique on a sample of \textit{Kepler} stars with small transiting planets and found that the planet-hosting stars have systematically higher values of $v\sin{i}$ than the control sample. Their control sample of stars, however, came from heterogeneous sources. \cite{Louden_2021} built upon the 2017 work using a carefully curated control sample that was observed and analyzed using the same pipelines as the \textit{Kepler} planet-hosting stars. The results showed a similar tendency for the average $v\sin i$ of the planet hosts to be larger than that of the control sample that is randomly oriented. \cite{2015ApJ...801....3M} also found evidence for misalignment for systems with hot stars and alignment for cool hosts using the distribution of observed rotational amplitudes, as this distribution depends on their projected axes of rotation. 

The effective temperature of the stars with misaligned planets spans the approximate value of the Kraft break, 6250 \,K \citep{kraft_studies_1967}. The Kraft Break delineates a threshold in stellar rotation: stars above 1.3 solar masses rotate relatively quickly, while those below this mass threshold exhibit significantly slower rotation rates. This change is primarily attributed to magnetic braking, a process where a star loses angular momentum through its stellar wind interacting with its magnetic field. This is related to the depth of the convective envelope of the star. The break typically occurs at temperatures around 6250 K, and stars on either side encompass F and G spectral-type stars. Dividing the planet-hosting star $v\sin i$ sample of \cite{Louden_2021} above and below this break revealed that planet-hosting stars with $\teff > 6250$ K tend to have a broader obliquity distribution (lower $ \langle v\sin{i} \rangle$ relative to the control stars) than planet-hosting stars with $\teff < 6250$ K. These results suggest that planets around hot stars are misaligned regardless of size and multiplicity. Observations of K2-290 via RM \citep{2019MNRAS.484.3522H} and Kepler-56\footnote{This star has an effective temperature below the Kraft Break now, but its mass is 1.3 $M_{\odot}$, so it was formerly a hot star before leaving the main sequence}  via helioseismology \citep{2013Sci...342..331H} are consistent with this conclusion. However, both the control and planet-hosting samples used by \cite{Louden_2021} were limited in size. In particular, hotter stars ($\teff > 6250$ K) were not well represented in the {\it Kepler} sample of planet-hosting stars and were limited in the control sample.  

Motivated by a desire to break the degeneracy between hot stars and hot Jupiters as the source of misalignment (high obliquity) and to study stellar obliquity for different types of planets around all types of stars, this study expands both the planet-hosting and control samples in the 6250-7100K range beyond those used by \cite{Louden_2021}, which had host stars with $5950 < \teff < 6550$ K. We include planet candidates from the NASA {\em TESS} mission \citep{2015JATIS...1a4003R} and a control sample of stars selected from the \textit{Gaia} \citep{2018A&A...616A...1G} catalog. This approach allows for a more comprehensive exploration of the higher effective temperature regime and its relationship to stellar obliquity, especially in stars hosting smaller planets, using a homogeneous data set.

This paper is organized as follows: Section \ref{sec:sample} outlines the analysis of the spectra to determine the spectroscopic parameters of the observed stars and the quality control methodology employed. Section \ref{sec:obliquities} analyzes the obliquity distributions of the planet-hosting and control stars. Finally, Section \ref{sec:summary} states the main conclusions of the paper, discusses the assumptions of the work, and puts the findings in context with current theories of stellar obliquities. Appendix \ref{sec:obs} describes the observational sample design and tests done to ensure homogeneity of the 2022-era samples.

\section{Observations and Sample Construction}\label{sec:sample}
The full details of the observations and the reductions can be found in Appendix \ref{sec:obs}. We provide a brief summary here.  

\subsection{Initial Target List}
Motivated by the evidence of misalignment (high obliquity, low $\langle v\sin{i}\rangle$) for hot stars presented in \cite{Louden_2021}, we used the TESS catalog to select planet hosts with $5950 < \teff < 7150 $ K and the Gaia catalog to choose control stars that were spectroscopically indistinguishable from the TESS planet hosts (all the effective temperatures came from Gaia broadband colors). The Gaia sample was magnitude-limited (limit of $G < 11$). We established indistinguishability via rejection sampling based on the magnitudes and effective temperatures. We further limited the targets by declination, Gaia radii, and flags for false alarms, false positives, and brown dwarf candidates on the TESS stars, resulting in 445 control stars and 59 planet-hosting stars.

\subsection{Observation, Stellar Parameters, and Updated Sample}
Of those, we observed 427 control stars and 57 planet-hosting stars with Keck/HIRES during the spring of 2022. The observations were spread over several months, amounting to two nights of Keck time. We used the same instrumental setup, observing protocols, data reduction software, and analysis procedures used by the CKS \citep{2017AJ....154..107P}. Following observations and data reduction, we utilized SpecMatch \citep{2015PhDT........82P} to determine our sample's spectroscopic parameters, including effective temperature ($T_{\rm eff}$), surface gravity ($\log g$), iron metallicity ([Fe/H]), and projected rotation velocity ($v\sin i$). As detailed in \cite{2015PhDT........82P}, SpecMatch provides these parameters with high precision: $\pm 60 K$ for $T_{\rm eff}$, $\pm 0.10$ dex for $\log g$, $\pm 0.04$ dex for [Fe/H], and $\pm 1.0 {\rm km,s^{-1}}$ for $v\sin i$. Using \texttt{isoclassify} \citep{2017ApJ...844..102H, 2020AJ....160..108B}, we also calculated the ages of the stars based on the spectroscopic parameters. We selected our sample by applying the following criteria:

\begin{eqnarray*}
5950\,{\rm K} < &T_{\rm eff}& < 7150\,{\rm K},\\
3.95 < &\log g& < 4.45,\\
-0.3 < &{\rm [Fe/H]}& < 0.3,\\
1.0 < &{R_{\odot}}& < 2.5 \\
\end{eqnarray*}

Following the same quality control procedures as \citet{FultonPetigura2018}, we also eliminated from consideration any star for which the \citet{Gaia2018} geometric parallax is not available or has a precision lower than 10\% (the typical level of uncertainty for stars in the relevant magnitude range is 5\%).

After down-selecting, our new 2022-era sample contained 20 planet hosts and 293 control stars. 


\subsection{Combined Sample Properties}

In the present paper, we expand our sample from that employed by \cite{Louden_2021} by incorporating data from the 2022-era observing runs, increasing the number and range of effective temperatures, especially for hot stars above 6250 K (24 control stars had $\teff > 6250$ in the 2021-era sample compared to 169 in the 2022-era sample). This extension allows for a more comprehensive analysis and clearer insight into the behavior across the entire temperature spectrum, particularly around the Kraft Break. 

\begin{figure*}
    \centering
    \includegraphics[width=0.9\textwidth]{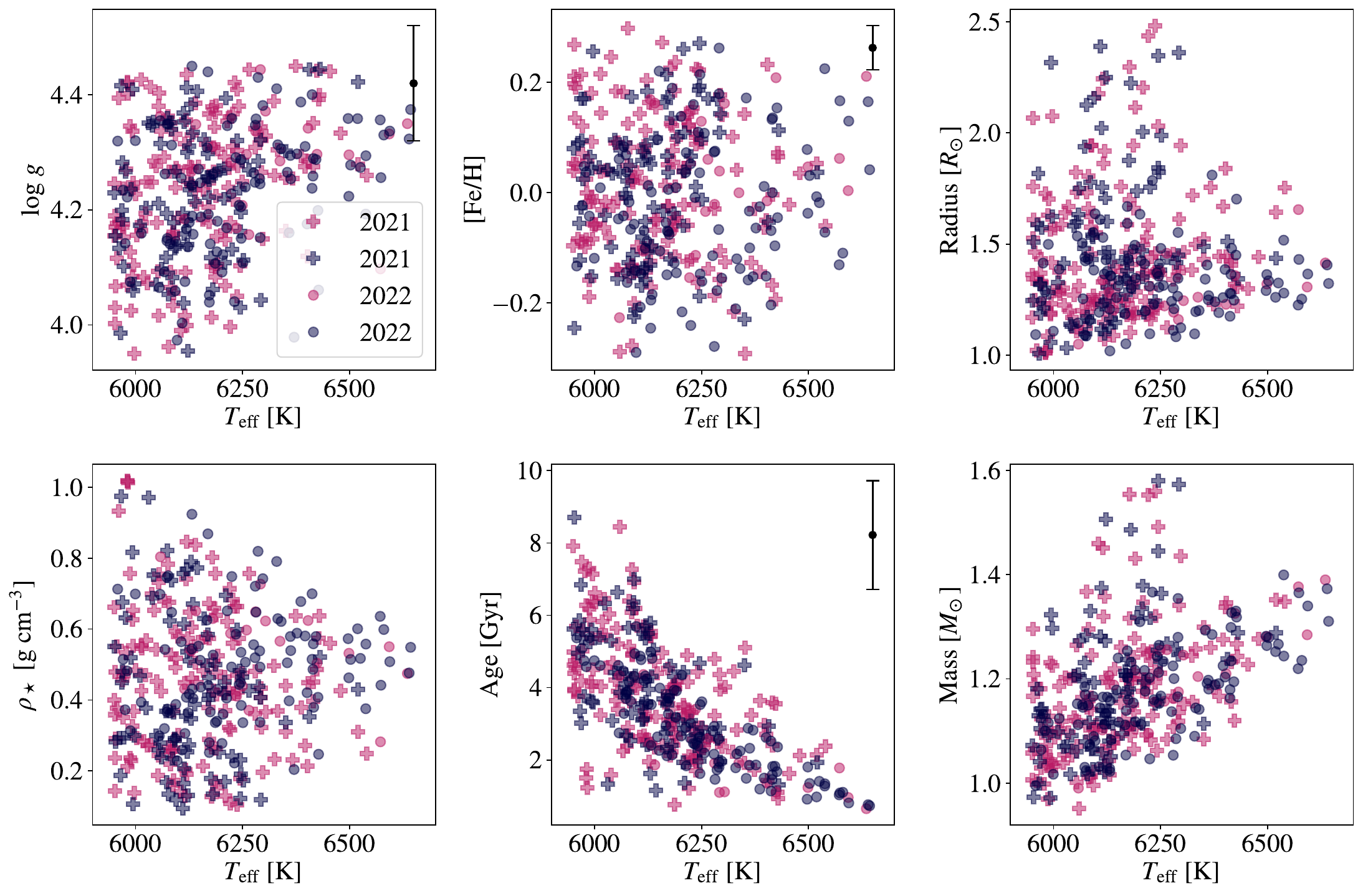}
    \caption{\textbf{Comparison between the properties of the control stars (navy) and the planet hosts (magenta) of both the 2021 (crosses) and 2022 data (dots).} The precision of the measurements is 60~K in $T_{\rm eff}$, 0.10~dex in $\log g$, 0.04~dex in [Fe/H], and 1.0 km\,s$^{-1}$ in $v\sin i$. The ages are determined by \texttt{isoclassify} using the MIST isochrones and have a typical precision of 30\%.}
    \label{fig:specparams}
\end{figure*}

The achieved goal of the observations was to obtain measurements for more control stars and planet hosts in the $\teff > 6250$ \,K range. Because of the limited TESS sample, there were many more control stars in this regime than planet hosts. Thus, we performed rejection sampling on the control stars so that the temperature, $\log g$, and [Fe/H] distributions were indistinguishable from those of the planet-hosting stars. While this cut the number of stars used in the final analysis, it is critical for the samples to be drawn from the same spectroscopic distributions. Initially, we derived probability density functions (PDFs) for both populations using Kernel Density Estimation (KDE). This non-parametric way of estimating the PDF allows for flexibility in capturing the distribution of stellar effective temperatures without assuming a predefined form. To conduct rejection sampling, we established a constant ensuring the scaled PDF of control stars always enveloped that of the planet hosts. The rejection sampling algorithm iteratively generated random numbers proportional to the value of the control stars' PDF, scaled by C, and accepted samples where these random numbers fell below the PDF of the planet-hosting stars. This method facilitated the generation of a subset of control stars whose effective temperature distribution closely mirrors that of the planet hosts, thereby allowing for a controlled comparison. The resultant sampled distribution of control stars shows the efficacy of rejection sampling in matching the statistical properties of these distinct stellar populations.

The final combined sample had 180 control stars and 170 planet-hosting stars. To confirm the null hypothesis that the planet hosts and control stars have indistinguishable distributions of spectroscopic parameters, we calculated the Anderson-Darling $p$-value and performed the two-sided Kolmogorov-Smirnoff test. For all cases, the $p$-value is much greater than 0.05 (see Table \ref{tab:pvaluesfullsample}).

We also performed the generalized version of the KS-test that checks for differences in the joint distribution of parameters \citep{PressTeukolsky1988, FasanoFranceschini1987, Peacock1983}. We found $p > 0.05$ for $T_{\rm eff}$ vs. $\log{g}$, $T_{\rm eff}$ vs. [Fe/H] and [Fe/H] vs. $\log{g}$. While these tests cannot guarantee a completely homogeneous sample, they tell us that the null hypothesis cannot be ruled out (which can also be seen in Figure \ref{fig:specparams}).

\begin{table}
    \centering
    \caption{\textbf{Statistics on spectroscopic parameters of combined 2021-era and 2022-era samples.}}
    \begin{tabular}{c|c|c}
        \hline
        Parameter & Anderson-Darling &  KS\\
        \hline
        $T_{\rm eff}$ & 0.145 & 0.179 \\
        \hline
        $\log{g}$ & 0.250 & 0.326 \\
        \hline
        [Fe/H] & 0.162 & 0.465 \\
        \hline
    \end{tabular}
    \label{tab:pvaluesfullsample}
\end{table}

\section{Hot Stars and High Obliquities} \label{sec:obliquities}

\subsection{Model Independent Test}
\begin{figure*}
    \centering
\includegraphics[width=\textwidth]{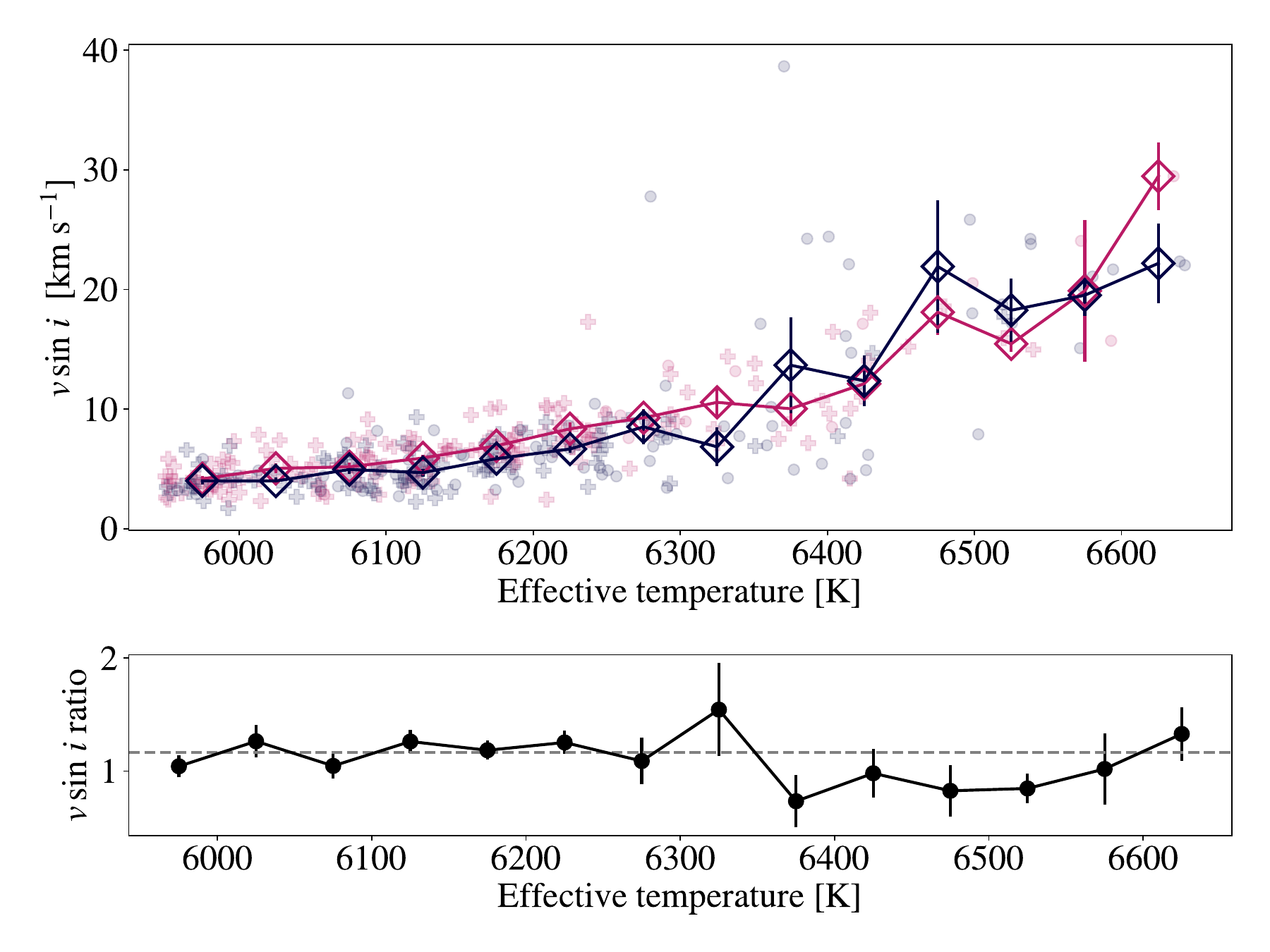}
    \caption{\textbf{Projected rotation velocity versus effective temperature.} The planet hosts (magenta) have slightly higher mean $v\sin{i}$ than the control stars (navy) below 6250 K, in agreement with \cite{Louden_2021} and indicating a tendency for spin-orbit alignment. The stars are binned in 50K intervals, and the diamonds mark the binned values of $v\sin{i}$. The bottom panel shows the ratio of $v\sin i$ of the planet hosts to the control stars for each bin with the requisite uncertainties. The dashed line shows the mean value of the ratio for stars below 6300 K. Every bin hotter than 6350 K is considerably below the value for cool stars, except for the hottest bin, which has a large error bar due to high scatter. The error bar for the final bin is calculated by estimating using the typical scatter in neighboring bins due to limited points in this bin.}
    \label{fig:vsiniteff}
\end{figure*}

Figure \ref{fig:vsiniteff} shows the projected rotation velocity versus the effective temperature of the stars from both the 2021-era and 2022-era samples. The effective temperatures are grouped into 50\,K bins, and each bin's average temperature is plotted. The planet hosts and control stars follow the expected upward trend in $v\sin{i}$ as temperature increases. 

The averages of the temperature bins for the planet hosts appear to be slightly higher than the control stars up to 6250\,K. Higher $v\sin{i}$ (lower obliquity) indicates a greater tendency for spin-orbit alignment. 

\begin{figure}
    \centering
\includegraphics[width=0.45\textwidth]{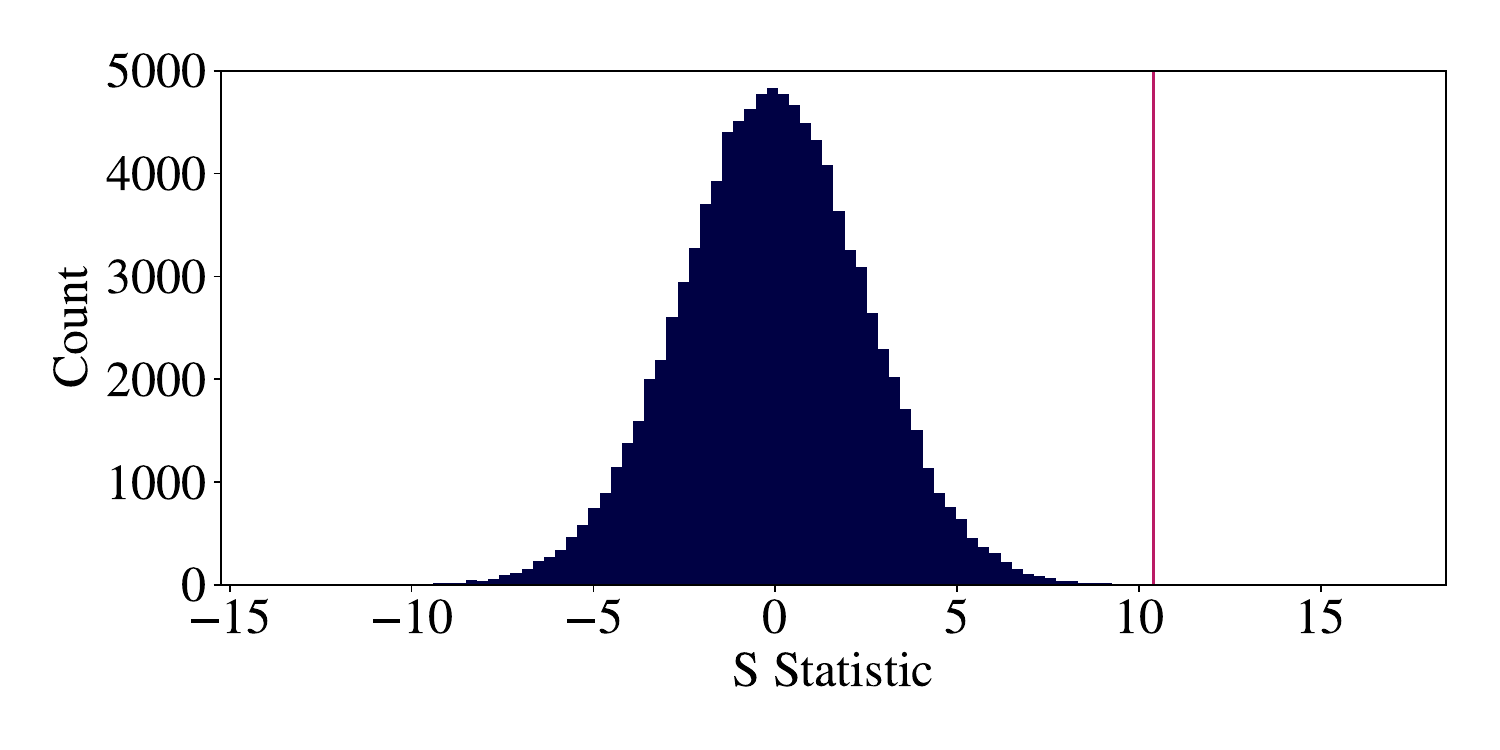}
    \caption{\textbf{Histogram of the summed deviation statistic (S) over 100,000 simulated trials.} The vertical line represents the observed value \( S_{\rm obs} = 10.41 \) in the real data. The histogram illustrates the distribution of \( S \) values across the simulated datasets, comparing planet-hosting and control stars below 6250 K. The extremely low occurrence of simulations exceeding the observed \( S_{\rm obs} \) value (35 in this case) indicates a statistically significant difference between the two groups, with a calculated p-value of 0.00035. This result strongly suggests that the observed differences in \( v\sin i \) distributions are not due to random chance. S is similar to a $\chi^2$ statistic, see Equation \ref{eq:S}.}
    \label{fig:S_stat}
\end{figure}

We quantify the difference in planet-hosting stars and control stars by asking how often differences at the level seen in Figure \ref{fig:vsiniteff} would occur by chance if $T_{\rm eff}$ and $v\sin i$ for all the stars were drawn from the same two-dimensional distribution. We created the simulated data sets by combining the planet hosts and control stars and then randomly drawing members of the combined sample to serve as ``planet hosts'' and as ``control stars'' in proportion to how many of each population are in the observed data. We quantified the difference between the two distributions with the statistic

\begin{equation}\label{eq:S}
   S \equiv
    \sum_{n=1}^8
    \frac{
      \langle v\sin i\rangle_{{\rm p},n} - \langle v\sin i\rangle_{{\rm c},n}
    }{
      \sqrt{ \sigma^2_{{\rm p},n} + \sigma^2_{{\rm c},n} }
    },
\end{equation}

where $\langle v\sin i\rangle_n$ is the mean value $v\sin i$ within the $n$th temperature bin; $\sigma_n$ is the corresponding standard deviation of the mean; and ``p'' and ''c'' refer to the planet and control samples, respectively. The real data have $S_{\rm obs} = 10.41$. After computing the $S$ statistic for $10^5$ simulated data sets, 35 had $S > S_{\rm obs}$, indicating  $p=0.00035$ and rejecting the null hypothesis that these two samples came from the same distribution (see Figure \ref{fig:S_stat}). 

\subsection{A Dual Model}
To further constrain the obliquity distribution of the planet-hosting stars, we apply the same dual modeling as in \cite{Louden_2021}. The model's foundation is that a randomly oriented sample of stars would have $\langle \sin i \rangle = \pi/4 \approx 0.785$. At the same time, if spins and orbits are always aligned, the spin-orbit-aligned transiting planet population would be expected to have $\langle \sin i \rangle = 1 $ (lower obliquities). Therefore, by constraining the value of $\langle \sin i \rangle$, we can assess the obliquity distribution of planet-hosting stars. All of the stars, planet hosts and control stars, were fit to the same model.

In this model,
the mean rotation velocity is
    \begin{equation}
     \langle v\rangle(\tau) = c_0 + c_1 \tau + c_2 \tau^2,
    \end{equation}
where
\begin{equation}
     \tau \equiv \frac{T_{\rm eff} - 6250\,{\rm K}}{300\,{\rm K}}
\end{equation}
varies from $-1$ to $+1$, and $c_0$, $c_1$, and $c_2$ are free parameters.
The mean $v\sin i$ value in the model depends
on whether the star is a control star or a planet host:
\begin{eqnarray}
\langle v\sin i\rangle_n &=& \langle v\rangle_n \times \frac{\pi}{4}~~({\rm control~stars}) \\
\langle v\sin i\rangle_n &=& \langle v\rangle_n \times \langle \sin i\rangle~~({\rm planet~hosts}),
\end{eqnarray}
where we have used the fact that $v$ and $\sin i$ are
uncorrelated.
Thus, in this model, the polynomial coefficients are constrained
by all of the stars, and the $\langle \sin i\rangle$ parameter
is constrained by the planet hosts.

The goodness-of-fit statistic was taken to be
\begin{equation}
    \chi^2 = \sum_{n=1}^{251}
    \left(
    \frac{ v\sin i_{{\rm obs}, n} - \langle v\sin i\rangle_{{\rm calc}, n} }
         {1\,{\rm km\,s}^{-1}}
    \right)^2,
\end{equation}
where $v\sin i_{{\rm obs}, j}$ is the observed value of $v\sin i$ of the $n$th star,
$\langle v\sin i\rangle_{{\rm calc}, i}$ is the mean value of $v\sin i$ calculated
according to the model,
and $1\,{\rm km\,s}^{-1}$ is the measurement uncertainty.

We list the premises of the modeling here:

\begin{enumerate}
    \item A star's rotation velocity $v$ and inclination $i$ are independent variables. 

    \item For any value of the effective temperature, the control stars and the planet hosts have the same distribution of rotation velocities (see the sample construction and comparisons presented in Section \ref{sec:sample}).

    \item The mean rotation velocity $\langle v\rangle$ is a quadratic function of effective temperature. Based on Figure \ref{fig:vsiniteff}.

    \item The measurements of $v\sin i$ for the control stars and the planet hosts are subject to the same systematic uncertainties.  Ensuring this is the case was the motivation for obtaining all the spectra with the same instrument and analyzing them with the same code.

    \item The control stars are randomly oriented in space.
\end{enumerate}
To these, we add a sixth premise based on the presence of the Kraft Break and consider two different cases:
\begin{itemize}
    \item[6a.] The obliquities of the transiting planet hosts
    are all drawn from the same distribution.
\setcounter{enumi}{4}
    \item[6b.] There are two different obliquity distributions:
    one for hosts cooler than 6250\,K,
    and one for hosts hotter than 6250\,K.
\end{itemize}

Because these data have uncertainties due to unknown distributions of intrinsic rotation velocity $v$ and $\sin{i}$, we used bootstrap resampling to establish confidence levels in the results. To sample from the joint probability density of the parameter values, we created $10^5$ simulated draws. Each draw has the same number of planet hosts and control stars as the actual data set. The model was fitted to each simulated data set by minimizing the $\chi^2$ statistic. 

We fit first to a single obliquity distribution and then divide the data into a set of distributions split at 6250\,K, motivated by the Kraft Break \citep{1930ApJ....72..267S,1967ApJ...150..551K}. This fitting is performed for the entire sample, the stars known to have more than one transiting planet (multis), and the sample cut in period and radius space. The full list of $\langle \sin i\rangle$ values from single and dual modeling of temperature ranges is contained in Table \ref{tab:sigmasandsinis}.

\subsection{All Planet Hosting Systems}

When fitting all planet-hosting systems to a single obliquity distribution, the bootstrap procedure gave $\langle \sin i\rangle = 0.83\pm 0.044$, $3.8 \sigma$ from alignment (low obliquities) and $1.0 \sigma$ from random orientations (broader obliquity distribution).

Motivated by the Kraft Break, we separated the planet-hosting stars into distributions above and below $6250\,K$. In this case, the hotter sample $\langle \sin i\rangle = 0.79 \pm 0.05$ is within $1 \sigma$ of random orientations (broader obliquity distribution). The cooler sample has a higher $\langle \sin i\rangle = 0.91 \pm 0.05$ (lower obliquities) and differs by $2.7 \sigma$ from random orientation.   

\subsection{Compact Multiplanet Systems}

We extend our analysis to examine the orbital alignment in multiplanet systems, noting that we cannot compare singles and multis, as many systems with only one planet detected may have more. Instead, we compare the subset of planets hosts with multiple known transiting planets in the inner system ($P < 100$  days) and period ratios less than six\footnote{The periods were from Exoplanet Archive data \citep{ps} accessed December 2023. This dataset or service is made available by the NASA Exoplanet Science Institute at IPAC, which is operated by the California Institute of Technology under contract with the National Aeronautics and Space Administration.} to the whole sample. A system was classified as multiplanet even if one or more companions were planet candidates. 

For stars with  $\teff > 6250 K$,  we find $\langle \sin i\rangle = 0.79 \pm 0.04$. This is $5.2 \sigma$ from perfect alignment (lower obliquities) and consistent at $0.04 \sigma$  with random orientation (higher obliquities). On the other hand, systems orbiting stars with $ \teff < 6250 K$ had $\langle \sin i\rangle$ of $0.89 \pm 0.04$, which is $2.7 \sigma$ from perfect alignment (lower obliquities). In the full temperature range of 5950 K to 7150 K, the systems had $\langle \sin i\rangle$ of $0.83 \pm 0.05$, consistent within $0.9 \sigma$ with random orientations (higher obliquities). 

\subsection{Splitting the Sample by Planet Radius and Period}
\begin{figure*}
    \centering
    \includegraphics[width=\textwidth]{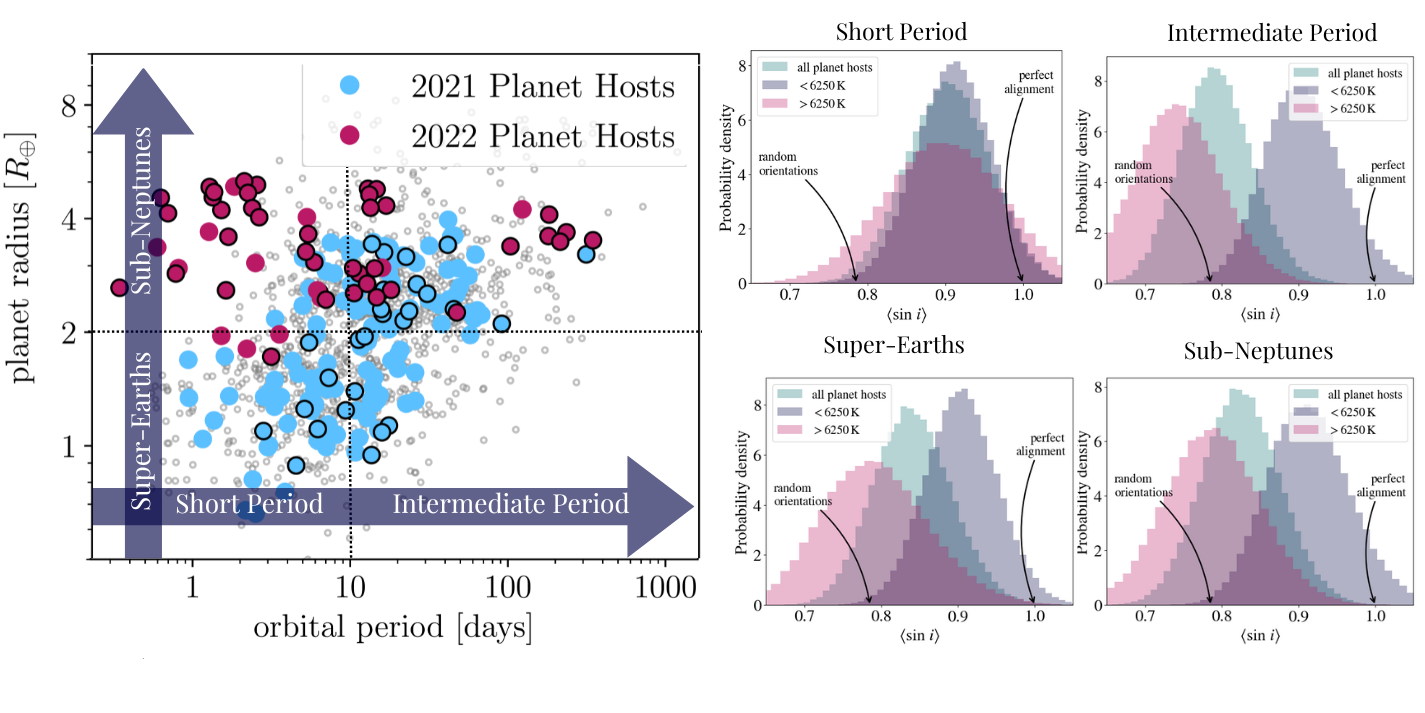}
    \caption{\textbf{Planetary properties for the combined sample used for analysis and probability density distribution split by planet radius or period.} \textit{Left:}  Planetary radius and orbital period for all the known transiting planets associated with the 153 planet hosts in \cite{Louden_2021} and the 20 planet hosts in the 2022 sample are shown on top of the full KOI sample. The planets outlined in black have hosts with $\teff > 6250$ K. \textit{Right:} We divided the planets into subsets based on radius and period and fit the sub-samples. The dashed lines on the left panel show the locations of the cuts over the planet sample in period-radius space. After performing each of the two cuts, the resulting distributions of $\langle \sin i\rangle$ are labeled according to the half they represent. }
    \label{fig:Divisions}
\end{figure*}

We further extended the analysis by examining the sample in planet period-radius space and asking whether planet radius or period has a noticeable impact on (mis)alignment. We divided the sample above and below $2$ Earth radii (super-Earths and sub-Neptunes) and above and below a 10-day period to split the sample into roughly equal subsamples (see Figure \ref{fig:Divisions}). 

Consistent among the results of both splits is that the hotter stars have lower $\langle \sin i\rangle$ (higher obliquities) than the cooler stars, except for the case of the hot planet-hosting stars ($P<10$ days). We find an approximately $1\sigma$ difference between the short and intermediate period planet-hosting stars:  $P<10 :\langle \sin i\rangle = 0.90 \pm 0.053, P>10 \langle \sin i\rangle = 0.79\pm0.047$. The short-period planet-hosting stars have a stronger tendency toward alignment. In contrast, the intermediate period planet-hosting stars show more distinction between their Kraft-break subdivided populations and thus have posteriors spanning a larger $\langle \sin i\rangle$ range. 

No statistically significant distinctions exist between the subdivided samples in radius space. The sub-Neptunes have $\langle \sin i\rangle = 0.83 \pm 0.05$ and the Super-Earths have $\langle \sin i\rangle = 0.84\pm 0.05$

\begin{table*}
\centering
\caption{\textbf{Average $\langle \sin i\rangle$ values and deviations from alignment (low obliquity) and random orientations (high obliquity).}}
\begin{tabular}{lccc}
\midrule
\multicolumn{4}{c}{\textbf{All Planet Hosting Systems}} \\
 & {$\langle \sin i\rangle$} & {$\sigma$ from alignment (1)} & {$\sigma$ from random orientation (0.785)}  \\
$\teff > 6250 $ & $0.79\pm 0.054$ & 3.9 & 0.09\\
$\teff < 6250 $ & $0.91\pm 0.046$ & 2.0 & 2.7\\
$5950 < \teff < 7150 $ & $0.83\pm 0.044$ & 3.8 & 1.0\\
\midrule
\multicolumn{4}{c}{\textbf{Multiplanet Systems}} \\
 & {$\langle \sin i\rangle$} & {$\sigma$ from alignment (1)} & {$\sigma$ from random orientation (0.785)}  \\
$\teff > 6250 $ & $0.79\pm 0.041$ & 5.2 & 0.04 \\
$\teff < 6250 $ & $0.89\pm 0.041$ & 2.7 & 2.6\\
$5950 < \teff < 7150 $ & $0.83\pm 0.046$ & 3.7 & 0.9 \\
\midrule
\multicolumn{4}{c}{\textbf{Short Period Planets $P < 10 $ Days}} \\
 & {$\langle \sin i\rangle$} & {$\sigma$ from alignment (1)} & {$\sigma$ from random orientation (0.785)}  \\
$\teff > 6250 $ & $0.90\pm 0.075$ & 1.4 & 1.5\\
$\teff < 6250 $ & $0.91\pm 0.049$ & 1.8 & 2.6\\
$5950 < \teff < 7150 $ & $0.90\pm 0.053$ & 1.8 & 2.2\\
\midrule
\multicolumn{4}{c}{\textbf{Intermediate Period Planets $P > 10 $ Days}} \\
 & {$\langle \sin i\rangle$} & {$\sigma$ from alignment (1)} & {$\sigma$ from random orientation (0.785)}  \\
$\teff > 6250 $ & $0.74\pm 0.057$ & 4.5 & 0.7\\
$\teff < 6250 $ & $0.91\pm 0.052$ & 1.8 & 2.4 \\
$5950 < \teff < 7150 $ & $0.79\pm 0.047$ & 4.4 & 0.2\\
\midrule
\multicolumn{4}{c}{\textbf{Sub-Neptunes $R_p > 2 R_{\oplus} $ }} \\
 & {$\langle \sin i\rangle$} & {$\sigma$ from alignment (1)} & {$\sigma$ from random orientation (0.785)}  \\
$\teff > 6250 $ & $0.79\pm 0.062$ & 3.4 & 0.06\\
$\teff < 6250 $ & $0.91\pm 0.054$ & 1.5 & 2.4 \\
$5950 < \teff < 7150 $ & $0.83\pm 0.050$ & 3.5 & 0.8\\
\midrule
\multicolumn{4}{c}{\textbf{Super-Earths $R_p < 2 R_{\oplus} $ }} \\
 & {$\langle \sin i\rangle$} & {$\sigma$ from alignment (1)} & {$\sigma$ from random orientation (0.785)}  \\
$\teff > 6250 $ & $0.79\pm 0.070$ & 3.0 & 0.12\\
$\teff < 6250 $ & $0.90\pm 0.047$ & 2.0 & 2.6 \\
$5950 < \teff < 7150 $ & $0.84\pm 0.051$ & 3.1 & 1.1 \\
\bottomrule
\end{tabular}

\label{tab:sigmasandsinis}
\end{table*}

\section{Discussion and Conclusions}\label{sec:summary}

We present results that suggest that it is hot stars, not hot Jupiters, that are the source of high obliquities.  We came to this conclusion by first establishing that TESS and \textit{Kepler} planet-hosting stars with $\teff$ between 5950 and 7150 K exhibit systematically higher $v\sin{i}$ values (lower obliquities) compared to control stars of indistinguishable spectral characteristics. The strongest signal that hot stars are the source of high obliquities is our evidence for a difference in the $\langle \sin i\rangle$ distributions for planet-hosting stars above and below 6250 K, with cooler stars yielding $\langle \sin i\rangle = 0.91 \pm 0.05$ (lower obliquities, nearer alignment) and hotter stars yielding $\langle \sin i\rangle = 0.79 \pm 0.05$ (higher obliquities, increased tendency for misalignment). 

We divided the systems by planet property and multiplicity and found a pattern of cooler hosts having higher $\langle \sin i\rangle$ (lower obliquities) and hotter hosts having  lower $\langle \sin i\rangle$ (higher obliquities) to be persistent in the data regardless of cuts made in radius space. 

This result that the sub-Neptunes and super-Earths have statistically similar obliquity distributions is consistent with recent theories on the co-formation of planets of this size. The consensus is that these planets (this radius and mass range) form from a single parent population and that envelope loss herds them into two populations \citep{2012ApJ...761...59L, 2013ApJ...775..105O, 2014ApJ...795...65J}. If this is correct, one wouldn't expect much of a difference in the obliquity distribution between these planet sizes. The fact that we don't see a difference in their distributions is consistent with this model.

We see differences in obliquity distributions based on the period of the planets at a $1\sigma$ level. However, the data sample is not large enough to probe differences in high and low-mass planets, with high statistical significance. Future RM measurements of small, typically compact planets, with either short periods ($P< 10$ days) or long periods ($P > 10$ days), are critically needed. These measurements, which are currently sparse, could provide further confirmation of the result of misalignment in hot stars. It is also of note that \cite{2015ApJ...801....3M} found that the low obliquity of planets around cool stars extends up to at least 50 days which is not consistent with our findings, further emphasizing the need to probe this parameter range on a system by system level. 

Our findings significantly extend the scope of previous studies (e.g., \cite{2017AJ....154..270W}, \cite{2018AJ....156..253M}, \cite{Louden_2021}) by incorporating a larger and hotter stellar sample, thereby offering a more comprehensive understanding of the relationship between stellar temperature, planetary characteristics, and stellar obliquity. We increased the number of control stars with $\teff$ above the Kraft Break by 500\% and the number of planet-hosting stars with $\teff$ above the Kraft Break by 50\%. This leads to a higher significance of the finding that hot stars with planets have higher obliquities. Using the increased homogeneous sample sizes, we also added new comparisons based on multiplicity, planet period, and planet radius, all done in pursuit of relating the demographic properties to the theories of planet formation and evolution that can lead to the observed misalignments.

Our study is grounded on the ability to observe the $v\sin{i}$ of stars with $5950 < T_{\rm eff} < 7150$ K. This regime spans the temperature range in which measurement of projected rotational velocities becomes observationally complicated due to fewer/wider spectral lines; line asymmetries due to oblateness, surface temperature variations, and stellar activity; instrumental limitations; and fast rotation that broadens spectral lines. To handle these factors, we made every effort to ensure that the assumption that the rotation velocities of the planet-hosts and control stars are drawn from the same distribution holds. By carefully matching spectroscopic parameters, performing observations using the same instrument, and reducing and analyzing all the data with the same pipeline, we minimize systematic differences to the greatest extent possible.

To contextualize our results within theoretical frameworks, we refer to \cite{albrecht_stellar_2022}.  They categorize processes influencing stellar obliquities into four broad categories. The first category is tidal interactions. An example scenario in this category is systems where the realignment timescale is shorter than the orbital decay timescale \citep{Hansen_2012, Valsecchi_2014}. Inertial waves can also realign the stars and planets. Created by tidal perturbations, a component of these waves drives the star toward $\Psi = 0~, 90~, \rm{or}~ 180^\circ$ \citep{Xue_2014, Li_2016}. If the outer zone of a star were decoupled from its interior, skin-deep realignment would also be possible \citep{2004ApJ...610..464D,2010ApJ...718L.145W}. This scenario is consistent with hot stars remaining misaligned because they rotate too quickly, and thus, their convective outer layers couple strongly to the interiors. 

The second category is primordial misalignment. These scenarios refer to stellar formation environments where the angular momentum of the disk and star are not aligned. In chaotic accretion, interactions between protostars can cause the accretion of some material to be oblique and result in a disk with a tilt or warp. Magnetic warping comes from a toroidal magnetic field that creates a Lorentz force and increases any existing misalignments \citep{2011MNRAS.412.2799F, 2011MNRAS.412.2790L}. Finally, resonances can generate misalignments between the disk and the star \citep{2013ApJ...778..169B, 2014MNRAS.440.3532L}. 

The third category is post-formation misalignment. In these scenarios, gravitational interactions between planets, disks, and the star shift the planet's orbital plane and can leave the system misaligned (high obliquity).

The final category is internal gravity waves. Under this scenario, the misalignment is completely independent of planet properties and instead is due to the random tumbling of the photospheres (internal gravity waves) of stars found in simulations of stars with convective cores and radiative envelopes \citep{2012ApJ...758L...6R, 2013ApJ...772...21R}. These internal gravity waves would only be expected in hot stars and thus are consistent with the findings here and in other papers that hot stars have broader obliquity distributions than cool stars. However, we find potential evidence for dependence on the planet period, which is not predicted by this theory.

\begin{figure*}
    \centering
    \includegraphics[width=\textwidth]{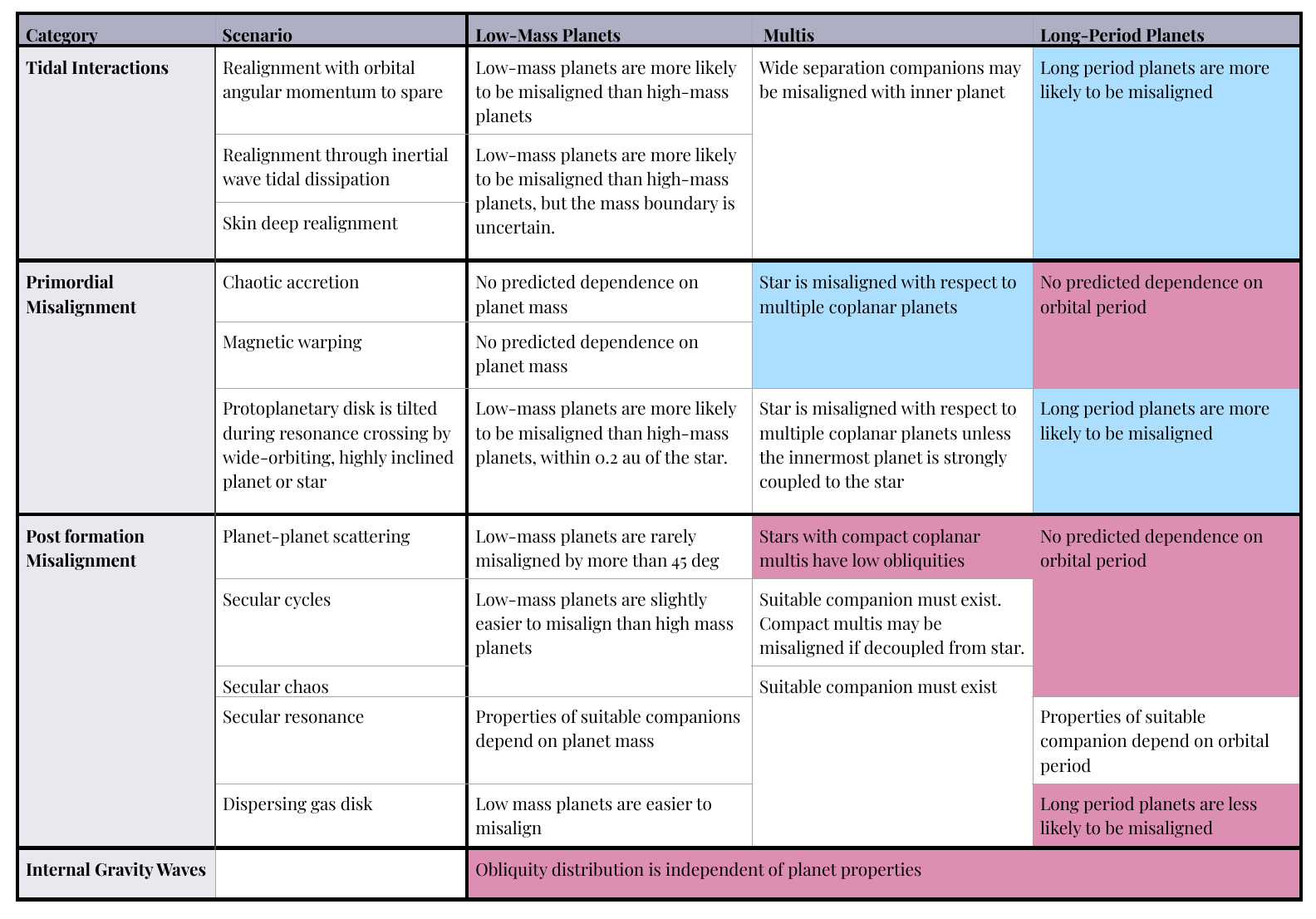}
    \caption{\textbf{Sources of the obliquity distribution and observational trends.} We reproduce selected columns of Table 3 of \cite{albrecht_stellar_2022}. Blue coloring indicates that the prediction is tentatively consistent with our data. Red coloring indicates that the observational data do not support that prediction.}
    \label{fig:Sources}
\end{figure*}

For each category, the theories predict trends for six different observational regimes (low-mass planets, multis, long-period planets, etc.). Figure \ref{fig:Sources} selects the columns that show the theoretical predictions for trends in the obliquities of three selected regimes that can be probed with data like that found in this paper: low-mass planets, multiplanet systems, and long-period planets. 

However, to make definitive statements on these regimes requires higher statistical precision than is possible with the current data. With the sample we currently have, the only observational parameters we can begin to make statements about are multiplicity and period length. With this, we color the multiplanet and long-period planet columns in Figure \ref{fig:Sources} based on (dis)agreement between observation and theory. The ubiquitous nature of misalignment and its relation to star and planet properties suggests that primordial misalignment or tidal interactions are more consistent with the broader demographic trends. However, based on the data, we cannot claim one category as the source of misalignment, nor would we expect to. Nature is under no obligation to have only one of these categories/scenarios be the source of the observed misalignment. In all likelihood, the observed trends come from a combination of processes. Nevertheless, it is useful to begin to discriminate among the most likely mechanisms using large demographic studies of obliquity, like that presented in this paper.

\begin{acknowledgments}
E.L. thanks Lily Zhao for helpful conversations during the observation proposal process, the Yale TAC for allocating time, and Garrett Levine, Sam Cabot, and Alex Polanski for insightful discussions about the project. 

S.W. is grateful for the support from the Heising-Simons Foundation (Grant No. 2023-4050) and the NASA Exoplanets Research Program NNH23ZDA001N-XRP (Grant No. 80NSSC24K0153). Thanks are also due to Xian-Yu Wang for valuable discussions on target selection.

J.M.A.M. is supported by the National Science Foundation Graduate Research Fellowship Program under Grant No. DGE-1842400.

We acknowledge the great cultural significance of the mountain of Maunakea to the indigenous Hawaiian community and are grateful for the opportunity to conduct observations from this Observatory. 

This work has made use of data from the European Space Agency mission Gaia (\url{https://www.cosmos. esa.int/gaia}), processed by the Gaia Data Processing and Analysis Consortium (DPAC,\url{https://www.cosmos.esa.int/ web/gaia/dpac/consortium}). Funding for the DPAC has been provided by national institutions, in particular, the institutions participating in the Gaia Multilateral Agreement.
\end{acknowledgments}
\facility{ADS, Exoplanet Archive}
\software{ 
\texttt{numpy} \cite{harris2020array};
\texttt{scipy} \cite{2020SciPy-NMeth};
\texttt{pandas} \cite{reback2020pandas}
\texttt{Astropy} \cite{2013A&A...558A..33A},
\texttt{Matplotlib }\cite{2007CSE.....9...90H}
}
\appendix 
\twocolumngrid
\section{Observations} \label{sec:obs}

\subsection{Target List Construction}
We started with the Gaia DR2 catalog to construct the observation target list. We selected stars with $6100 < T_{\textrm{eff}} < 7100$, based on Gaia broadband colors, Gaia g magnitude less than 11.3, declination between 20 and 89 $\deg$, and a Gaia-determined radius of $R_{*} <  2.5 R_{\odot}$. These constraints resulted in a list of 62577 stars. 

Next, we took the list of {\it TESS} stars not flagged as false alarms, false positives, or brown dwarfs, in the range $6100 < T_{\textrm{eff}} < 7100$ based on Gaia broadband colors, with a planet radius $R_p < 5 R_{\bigoplus}$, a stellar radius of $R_{*} <  2.5 R_{\odot}$ from TIC v8, and declination between 20 and 89 $\deg$ to be observable by \textit{Keck/HIRES}. This process resulted in a list of 59 targets. 

\begin{figure}
    \centering
    \includegraphics[width=0.45\textwidth]{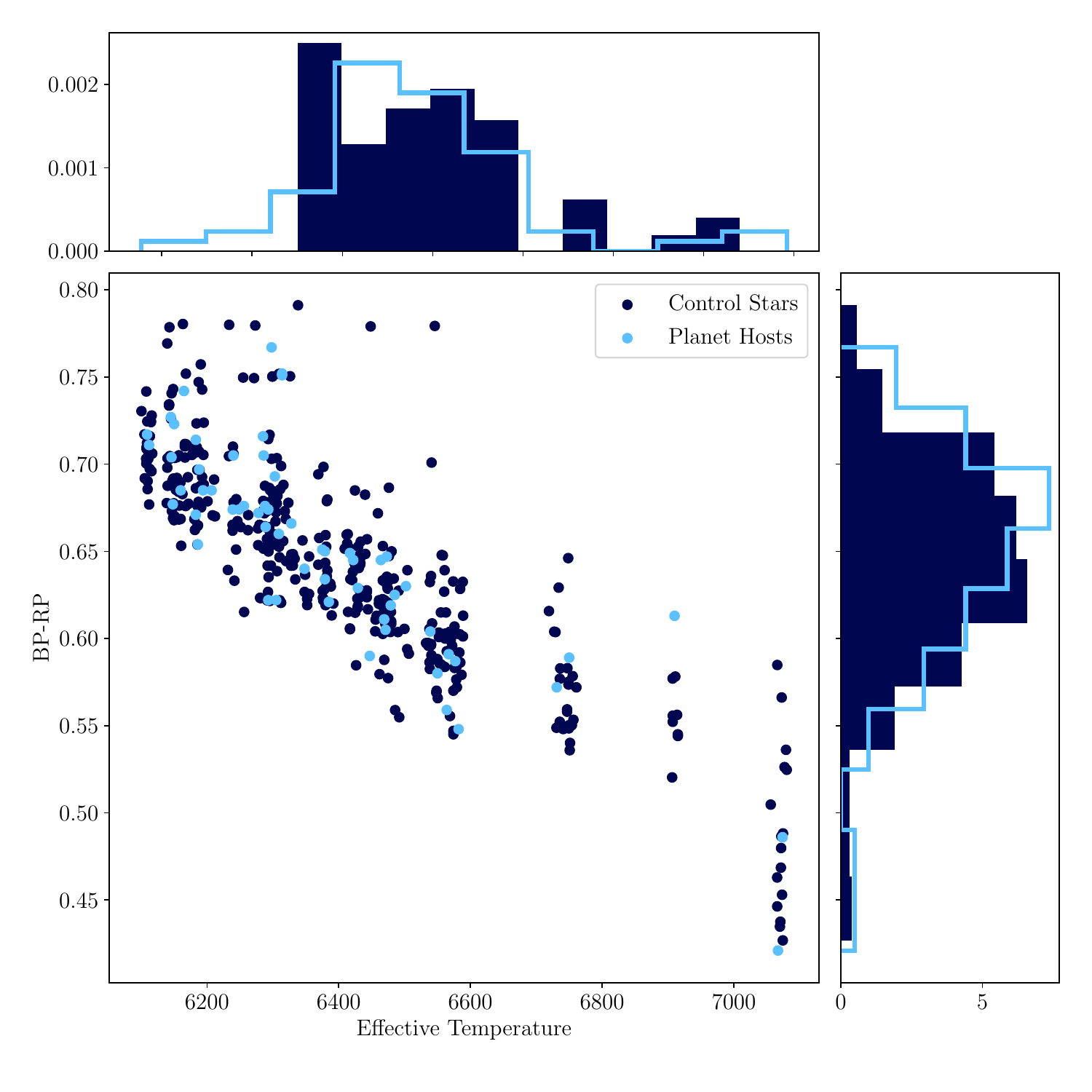}
   \caption{\textbf{Target list created using rejection sampling.} The striation is caused by using TESS exoFOP temperatures that are assigned. The stars span a temperature range of 6100-7100 K. }
    \label{fig:HRHist}
\end{figure}

To construct a control sample as similar as possible to the {\it TESS} planet hosts but selected without regard to rotation rate or orientation, we used rejection sampling to select a sample that is indistinguishable in Gaia BP-RP magnitude and effective temperature. We used the Gaia effective temperatures for both sets of stars. We estimated the pdf using kernel density estimation and then performed rejection sampling on the estimated distributions (see Figure \ref{fig:HRHist}). 

The temperatures of the planet hosts and control stars, as well as the BP-RP magnitudes, both pass the two-sided Kolmogorov Smirnoff test for indistinguishability with $p > \textrm{statistic}$. This rejection sampling resulted in 648 control stars and 59 planet hosts.

\begin{figure}
    \centering
    \includegraphics[width=0.45\textwidth]{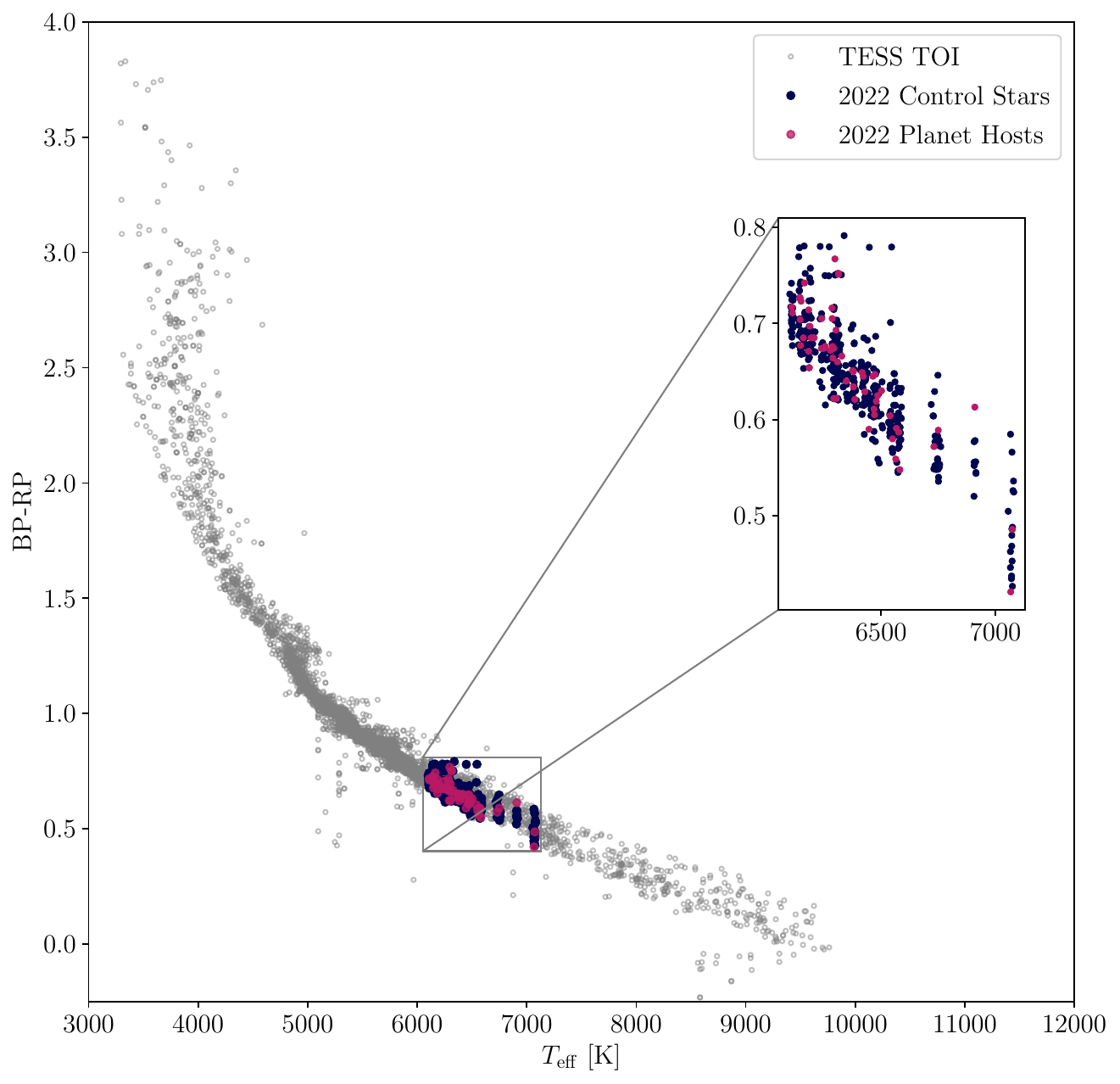}
   \caption{\textbf{Observational sample in the context of TESS TOIs.} The planet-hosting and control stars sample compared to the TESS TOI catalog. The striation is caused by using TESS exoFOP temperatures that are assigned. The stars span a temperature range of 6100-7100 K. }
    \label{fig:BPRPall}
\end{figure}

Next,  we reduced the number of control stars to fit into the two nights of observation time. Estimating exposure times and limiting the list resulted in 445 control stars. This limited sample also passed the two-sided Kolmogorov Smirnoff test when compared in BP-RP magnitude and effective temperature to the planet-hosting stars.

\begin{table}
    \centering
    \caption{\textbf{Statistics prior to observation.} Results of two-sided Kolmorgorov-Smirnoff tests on the planet-hosting and control stars after rejection sampling and trimming to fit the allotted observation schedule.}
    \begin{tabular}{c|c|c}
        \hline
        Parameter & P value &  Statistic\\
        \hline
        Magnitude & 0.52 & 0.11 \\
        \hline
        Temperature & 0.50 & 0.011 \\
        \hline
    \end{tabular}
    \label{tab:pvaluestargetlist2}
\end{table}

\subsection{2021 and 2022 Sample Comparisons}
\begin{figure}
    \centering
    \includegraphics[width=0.4\textwidth]{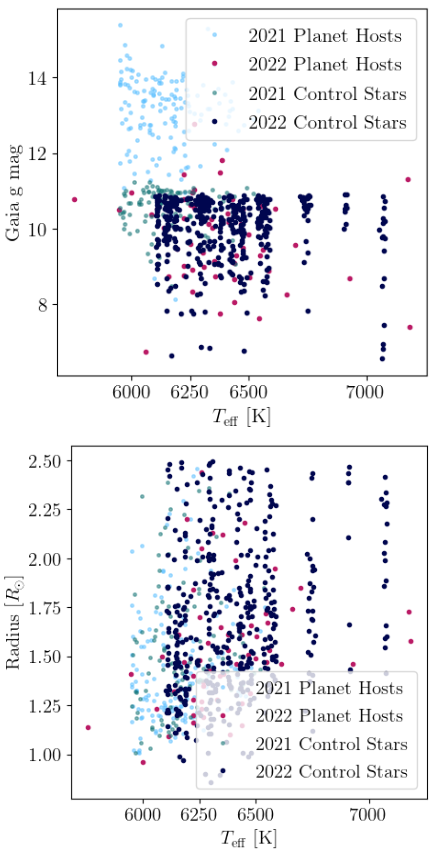}
    \caption{\textbf{Comparison of the 2021 and 2022 samples in effective temperature, Gaia g magnitude, and stellar radius.} The striations result from the exoFOP catalog used to make the observational target list.}
    \label{fig:compsamples}
\end{figure}

Compared to the 2021-era sample, this sample has slightly larger planet radii, as shown in Figure \ref{fig:Divisions}. Compared to the 2021-era sample, the overall magnitudes are brighter, enabling the collection of four times as many stars in only two nights of observations, as shown in Figure \ref{fig:compsamples}. 

\subsection{Reduction and Analysis}
To further ensure homogeneity between the samples, we observed all the stars on {\it Keck/HIRES} and used the same pipeline to analyze the spectra. Homogeneity in this process is necessary because measurements of $v\sin i$ are subject to systematic errors related to instrumental resolution and treatment of other line-broadening mechanisms. 

We observed 445 candidate control stars and 59 planet-hosting stars with {\it Keck/HIRES} during the spring of 2022. The observations were spread over several months, amounting to two nights of Keck time. We used the same instrumental setup, observing protocols, data reduction software, and analysis procedures used by the CKS.

\subsection{Post-Observation Planet Host Sample}
While we observed 59 new planet hosts, this paper added only 20 planet-hosting stars to the catalog used for analysis. The stars were cut based on 1. contaminated spectra, 2. not fitting the spectroscopic criteria, and 3. because of tests for false positives in the planet-hosting sample.

These stars were chosen from the TESS candidate list to fit the necessary magnitude, temperature, and planet size constraints to be eligible for this study. The region where many of the new TESS candidates lie is near the hot Neptune desert, named as such because short-period small planets seem to be rare around hot stars \citep{2022ApJ...935L..10G}. Therefore necessarily, not all of these planets are confirmed (at the time of publication, four are known planets, and two are confirmed planets). We removed any planets from the work that were deemed false positives after the observational sample was created and the data was collected and before submission. However, in all likelihood, there are still false positive planets persistent in the data. 

There are 276 planets in the TESS planet with the same spectroscopic constraints as described in Section \ref{sec:sample}. Of the 276, 12 are known planets, 18 are confirmed planets, four are ambiguous planetary candidates, 186 are planetary candidates, and 45 have been found to be false positives. This suggests that historically 17\% of planets in this radius-period space have turned out to be false positives. Therefore, it is likely there is still some remaining impurity in the data. Despite this, our results robustly indicate a trend of higher obliquities in hotter stars.


\section{Samples}
\begin{deluxetable}{lcccc}
\tabletypesize{\scriptsize}
\tablecaption{Planet Hosting Stars 2022 Sample}
\tablehead{
\colhead{TIC} & \colhead{$T_{\textrm{eff}}$}& \colhead{$\log_{g}$} & \colhead{[Fe/H]} &
\colhead{$v\sin{i}$} \\
} 
\startdata 
 81212289 &          6292 &          4.44 &     0.02 &  13.63 \\
387603144 &          6114 &          4.32 &     0.00 &   5.38 \\
175193677 &          6141 &          4.35 &     0.15 &   6.53 \\
 75589027 &          6058 &          4.36 &    -0.24 &   2.85 \\
154568734 &          6180 &          4.31 &    -0.08 &   6.19 \\
165530380 &          6304 &          4.37 &     0.15 &   8.96 \\
284361752 &          6280 &          4.34 &    -0.04 &   8.86 \\
461387302 &          6389 &          4.44 &    -0.20 &   6.72 \\
232971294 &          6130 &          4.27 &     0.16 &   6.18 \\
120960812 &          6259 &          4.30 &    -0.05 &   9.47 \\
357972447 &          6303 &          4.31 &    -0.04 &  11.24 \\
148782377 &          6424 &          4.30 &     0.20 &  17.13 \\
349488688 &          6222 &          4.28 &    -0.17 &   7.59 \\
 43429656 &          6249 &          4.21 &     0.12 &   9.82 \\
237201858 &          6403 &          4.34 &    -0.03 &   8.50 \\
326114850 &          6156 &          4.14 &     0.12 &   4.70 \\
164458714 &          6337 &          4.28 &    -0.01 &  13.16 \\
277507814 &          6265 &          4.16 &     0.05 &   7.58 \\
158075010 &          6499 &          4.30 &     0.04 &  20.51 \\
 91987762 &          6236 &          4.17 &    -0.12 &   7.44 \\
283829553 &          6114 &          4.18 &    -0.11 &   6.74 \\
284564230 &          6491 &          4.22 &     0.17 &  28.75 \\
422838133 &          6438 &          4.27 &    -0.06 &   9.35 \\
230075120 &          6272 &          4.21 &    -0.08 &   9.95 \\
129539786 &          6479 &          4.35 &    -0.13 &  12.54 \\
219501568 &          6572 &          4.10 &     0.06 &  24.05 \\
358186451 &          6649 &          4.34 &    -0.05 &  21.56 \\
377290815 &          6191 &          4.09 &    -0.14 &   5.53 \\
233680651 &          6635 &          4.35 &     0.18 &  29.46 \\
233009109 &          6593 &          4.33 &    -0.01 &  15.71 \\
288471040 &          6239 &          4.06 &     0.16 &   9.17 \\
 17416749 &          6282 &          4.10 &    -0.09 &   8.68 \\
\enddata
\end{deluxetable}

\startlongtable
\begin{deluxetable}{lcccc}
\tabletypesize{\scriptsize}
\tablecaption{Control Stars 2022 Sample}
\tablehead{
\colhead{Star} & \colhead{$T_{\textrm{eff}}$}& \colhead{$\log_{g}$} & \colhead{[Fe/H]} &
\colhead{$v\sin{i}$} \\
} 
\startdata 
     ag+371739 &          6387 &          4.26 &     0.01 &   7.55 \\
     bd+212241 &          6073 &          4.35 &    -0.16 &   4.25 \\
     bd+214472 &          6242 &          4.21 &     0.10 &  10.44 \\
     bd+234821 &          6393 &          4.44 &     0.00 &  11.94 \\
     bd+242453 &          6470 &          4.32 &    -0.14 &   7.51 \\
     bd+244878 &          6172 &          4.04 &    -0.01 &   6.31 \\
       bd+2671 &          6438 &          4.36 &     0.02 &  27.32 \\
     bd+282405 &          6231 &          4.04 &     0.16 &   6.99 \\
     bd+283145 &          6457 &          4.32 &     0.08 &  14.88 \\
     bd+293160 &          6385 &          4.39 &    -0.00 &  19.05 \\
     bd+293565 &          6667 &          4.34 &     0.04 &  13.70 \\
     bd+324493 &          6538 &          4.29 &    -0.08 &  23.79 \\
      bd+32450 &          6330 &          4.41 &    -0.23 &   8.56 \\
     bd+332453 &          5971 &          4.23 &    -0.03 &   2.88 \\
     bd+361344 &          6536 &          4.34 &    -0.06 &  11.13 \\
     bd+362372 &          6204 &          4.12 &    -0.23 &   5.81 \\
     bd+372028 &          6076 &          4.35 &    -0.17 &   3.68 \\
     bd+372507 &          6276 &          4.26 &    -0.13 &   9.03 \\
     bd+382296 &          6406 &          4.26 &     0.10 &   9.14 \\
      bd+41234 &          6536 &          4.31 &     0.09 &  25.25 \\
     bd+414478 &          6373 &          4.33 &    -0.07 &  14.82 \\
     bd+422038 &          6399 &          4.39 &    -0.14 &  11.93 \\
     bd+422251 &          6225 &          4.36 &    -0.25 &   6.71 \\
     bd+432968 &          6676 &          4.36 &     0.06 &  11.15 \\
     bd+442433 &          6396 &          4.31 &     0.00 &   5.43 \\
     bd+452328 &          6069 &          4.26 &     0.08 &   5.29 \\
     bd+452420 &          6251 &          4.15 &    -0.12 &   5.06 \\
     bd+463014 &          6134 &          4.16 &    -0.19 &   3.25 \\
    bd+471622b &          6190 &          4.13 &    -0.00 &   3.92 \\
     bd+521550 &          6332 &          4.31 &    -0.13 &   4.23 \\
     bd+572357 &          6067 &          4.09 &    -0.04 &   3.55 \\
     bd+581829 &          6115 &          4.08 &    -0.08 &   5.77 \\
     bd+611764 &          6213 &          4.17 &     0.02 &   5.98 \\
     bd+612054 &          6151 &          4.25 &     0.21 &   6.21 \\
     bd+612187 &          6460 &          4.35 &     0.26 &  25.87 \\
      bd+61693 &          6131 &          4.24 &    -0.25 &   5.96 \\
    bd+631313b &          6291 &          4.24 &     0.17 &   3.42 \\
     bd+641869 &          6074 &          4.15 &     0.03 &  11.32 \\
     bd+651144 &          6297 &          4.38 &    -0.17 &   7.91 \\
      bd+67558 &          6283 &          4.37 &     0.02 &   6.90 \\
     bd+681326 &          6519 &          4.36 &    -0.02 &  18.74 \\
     bd+691033 &          6246 &          4.29 &     0.20 &   8.38 \\
      bd+69849 &          6466 &          4.31 &    -0.13 &   4.96 \\
      bd+76711 &          6214 &          4.35 &     0.06 &   6.47 \\
      bd+79384 &          6391 &          4.39 &    -0.03 &  13.86 \\
      bd+81129 &          6170 &          4.19 &    -0.08 &   5.58 \\
      bd+81206 &          6246 &          4.20 &     0.02 &   5.09 \\
       bd+8726 &          6363 &          4.33 &     0.05 &   5.56 \\
        118525 &          6447 &          4.28 &     0.08 &  16.98 \\
        123364 &          6377 &          4.30 &    -0.08 &   4.94 \\
         15267 &          6426 &          4.20 &     0.15 &   4.90 \\
         16351 &          6459 &          4.27 &    -0.02 &  19.10 \\
        192681 &          6247 &          4.15 &    -0.16 &   4.53 \\
        193388 &          6484 &          4.28 &     0.08 &  14.18 \\
        204069 &          6467 &          4.37 &     0.03 &   9.65 \\
        211973 &          6416 &          4.29 &    -0.01 &   4.18 \\
         22373 &          6257 &          4.24 &     0.10 &   4.38 \\
        227620 &          6355 &          4.27 &     0.16 &  17.12 \\
        234167 &          6354 &          4.40 &    -0.04 &  10.13 \\
        236520 &          6401 &          4.31 &     0.11 &   7.76 \\
        236616 &          6092 &          4.06 &    -0.06 &   4.57 \\
        238054 &          5988 &          4.07 &    -0.04 &   4.13 \\
        238279 &          6321 &          4.43 &    -0.15 &   6.13 \\
        240072 &          6454 &          4.42 &    -0.10 &  11.92 \\
        257207 &          6373 &          4.33 &     0.07 &  10.61 \\
          2601 &          6172 &          4.05 &    -0.08 &   5.81 \\
         29157 &          6515 &          4.31 &    -0.03 &   9.59 \\
         31016 &          6174 &          4.14 &    -0.09 &   3.25 \\
         33252 &          6286 &          4.43 &    -0.19 &   8.07 \\
        334074 &          6342 &          4.25 &     0.14 &   9.22 \\
        334836 &          6528 &          4.40 &     0.17 &  17.53 \\
        334875 &          6498 &          4.22 &    -0.04 &  18.00 \\
        335229 &          6413 &          4.40 &    -0.09 &  16.10 \\
        336166 &          6148 &          4.19 &    -0.25 &   6.02 \\
        336398 &          6344 &          4.31 &     0.02 &   7.86 \\
        338798 &          6530 &          4.31 &     0.16 &  16.51 \\
        344606 &          6622 &          4.44 &     0.09 &  24.22 \\
        345507 &          6281 &          4.36 &     0.06 &   6.87 \\
        347815 &          6422 &          4.29 &    -0.11 &  12.12 \\
          4697 &          6486 &          4.26 &     0.09 &  15.03 \\
         50226 &          6094 &          4.15 &    -0.16 &   8.18 \\
         66949 &          6081 &          4.14 &    -0.03 &   3.54 \\
         80887 &          6631 &          4.43 &    -0.04 &  15.91 \\
         84496 &          6396 &          4.36 &    -0.04 &   6.50 \\
         86414 &          6444 &          4.35 &     0.08 &  18.70 \\
         88436 &          6499 &          4.28 &     0.02 &   7.15 \\
         93090 &          6573 &          4.26 &     0.16 &  20.23 \\
         93899 &          6491 &          4.28 &    -0.10 &  17.50 \\
         94746 &          6374 &          4.22 &     0.16 &  29.88 \\
         96217 &          6509 &          4.17 &     0.01 &  12.59 \\
        ton935 &          6381 &          4.21 &     0.15 &  18.66 \\
 tyc1371-544-1 &          6363 &          4.25 &    -0.11 &   8.58 \\
 tyc1388-919-1 &          6147 &          4.27 &     0.09 &   3.69 \\
tyc1425-1013-1 &          6428 &          4.06 &    -0.05 &   6.17 \\
 tyc1483-493-1 &          6122 &          4.14 &    -0.10 &   4.00 \\
tyc1533-1075-1 &          6511 &          4.23 &     0.08 &  33.76 \\
 tyc1747-644-1 &          6433 &          4.02 &    -0.05 &  34.53 \\
  tyc1773-97-1 &          6340 &          4.31 &    -0.15 &   7.74 \\
 tyc1782-252-1 &          6290 &          4.25 &    -0.08 &  11.95 \\
 tyc1939-407-1 &          6503 &          4.28 &     0.02 &   7.89 \\
 tyc1958-390-1 &          6118 &          4.17 &    -0.10 &   4.31 \\
 tyc1963-457-1 &          6152 &          4.19 &    -0.11 &   6.10 \\
 tyc2014-615-1 &          6057 &          4.35 &     0.05 &   4.40 \\
tyc2032-1410-1 &          6279 &          4.25 &     0.06 &   5.66 \\
 tyc2057-575-1 &          6303 &          4.22 &    -0.09 &  11.12 \\
tyc2150-2294-1 &          6272 &          4.32 &     0.02 &   6.84 \\
tyc2205-2178-1 &          6497 &          4.36 &     0.07 &  25.83 \\
  tyc2224-99-1 &          6202 &          4.09 &     0.12 &   6.85 \\
 tyc2264-105-1 &          6363 &          4.14 &    -0.16 &  10.90 \\
tyc2308-1119-1 &          6109 &          4.00 &    -0.16 &   2.72 \\
tyc2310-1160-1 &          6421 &          4.38 &    -0.12 &  24.97 \\
 tyc2310-159-1 &          6643 &          4.37 &     0.03 &  22.02 \\
tyc2336-2248-1 &          6377 &          4.27 &    -0.20 &  10.22 \\
tyc2452-1483-1 &          6386 &          4.39 &    -0.04 &  24.23 \\
 tyc2459-419-1 &          6303 &          4.34 &     0.03 &   6.79 \\
tyc2462-1482-1 &          5980 &          4.22 &     0.03 &   3.78 \\
 tyc2494-801-1 &          6288 &          4.32 &    -0.04 &   9.55 \\
 tyc2505-219-1 &          6378 &          4.11 &    -0.00 &  37.22 \\
tyc2508-1246-1 &          6084 &          4.31 &    -0.08 &   3.17 \\
 tyc2523-360-1 &          6148 &          4.22 &    -0.15 &   6.14 \\
 tyc2543-380-1 &          6371 &          3.98 &    -0.16 &  38.64 \\
tyc2562-1079-1 &          6359 &          4.16 &    -0.17 &   6.97 \\
 tyc2566-225-1 &          6457 &          4.13 &    -0.25 &  31.65 \\
 tyc2586-210-1 &          6446 &          4.27 &    -0.20 &  21.53 \\
tyc2591-1206-1 &          6151 &          4.24 &     0.26 &   4.74 \\
 tyc2623-623-1 &          6519 &          4.30 &     0.10 &  21.84 \\
 tyc2681-855-1 &          6280 &          4.35 &    -0.29 &  27.76 \\
tyc2688-2670-1 &          6510 &          4.32 &     0.03 &  18.04 \\
tyc2733-1547-1 &          6340 &          4.21 &     0.03 &  11.50 \\
tyc2789-1572-1 &          6186 &          4.11 &    -0.14 &   4.55 \\
 tyc2799-295-1 &          6415 &          4.26 &     0.12 &  22.10 \\
tyc2808-1479-1 &          6525 &          4.31 &    -0.12 &  12.16 \\
tyc2809-1255-1 &          6344 &          4.44 &     0.20 &   9.36 \\
 tyc2814-376-1 &          6302 &          4.17 &    -0.03 &   8.50 \\
 tyc2982-972-1 &          6073 &          4.15 &     0.05 &   7.16 \\
 tyc2987-540-1 &          6335 &          4.19 &     0.27 &  10.76 \\
 tyc2988-757-1 &          6427 &          4.23 &    -0.07 &  14.50 \\
 tyc3038-803-1 &          6097 &          3.97 &    -0.30 &   4.81 \\
 tyc3079-318-1 &          6299 &          4.18 &     0.08 &  12.53 \\
tyc3126-2986-1 &          6389 &          4.13 &    -0.30 &  32.11 \\
tyc3131-1871-1 &          6317 &          4.32 &    -0.23 &   5.53 \\
 tyc3182-213-1 &          6430 &          4.31 &     0.26 &  15.67 \\
tyc3189-2000-1 &          6411 &          4.39 &    -0.16 &   8.40 \\
tyc3190-1103-1 &          6369 &          4.20 &     0.13 &  25.93 \\
 tyc3192-274-1 &          6572 &          4.36 &    -0.15 &  15.09 \\
 tyc3218-917-1 &          6731 &          4.38 &     0.13 &  20.30 \\
 tyc3237-736-1 &          6466 &          4.34 &     0.04 &  25.38 \\
 tyc3349-103-1 &          6445 &          4.22 &    -0.22 &  12.60 \\
tyc3353-1546-1 &          6275 &          4.43 &    -0.15 &   6.29 \\
 tyc3370-942-1 &          6473 &          4.36 &     0.01 &   9.01 \\
tyc3399-1474-1 &          6319 &          4.43 &     0.02 &  31.32 \\
tyc3401-1067-1 &          6525 &          4.27 &    -0.03 &  17.18 \\
 tyc3406-881-1 &          6342 &          4.40 &    -0.14 &  10.14 \\
tyc3410-1187-1 &          6412 &          4.24 &    -0.22 &   8.84 \\
tyc3413-2037-1 &          6159 &          3.97 &    -0.18 &   7.26 \\
 tyc3421-745-1 &          6362 &          4.20 &    -0.06 &  10.44 \\
 tyc3440-869-1 &          6068 &          4.35 &    -0.10 &   2.05 \\
 tyc3444-230-1 &          6133 &          4.19 &    -0.02 &   8.23 \\
tyc3452-1037-1 &          6013 &          4.16 &     0.00 &   4.87 \\
 tyc3458-338-1 &          5999 &          4.32 &    -0.13 &   4.45 \\
 tyc3513-864-1 &          6441 &          4.38 &    -0.14 &  22.67 \\
 tyc3544-488-1 &          6547 &          4.25 &    -0.07 &  18.73 \\
tyc3551-1800-1 &          6234 &          4.29 &    -0.02 &   5.11 \\
 tyc3553-865-1 &          6538 &          4.19 &     0.22 &  24.22 \\
tyc3566-1156-1 &          6374 &          4.27 &     0.01 &   7.53 \\
 tyc3568-313-1 &          6191 &          4.18 &    -0.16 &   8.26 \\
tyc3578-2501-1 &          6560 &          4.44 &     0.21 &  13.20 \\
tyc3578-2673-1 &          6640 &          4.32 &     0.16 &  22.33 \\
 tyc3771-651-1 &          6360 &          4.41 &     0.06 &  10.78 \\
tyc3808-1772-1 &          6107 &          4.26 &    -0.22 &   4.52 \\
tyc3815-1306-1 &          6143 &          4.07 &    -0.08 &   7.13 \\
 tyc3818-445-1 &          6416 &          4.18 &     0.13 &  14.70 \\
 tyc3821-633-1 &          6362 &          4.29 &    -0.05 &  10.17 \\
  tyc3829-95-1 &          6436 &          4.20 &     0.01 &  16.30 \\
 tyc3836-870-1 &          6055 &          4.06 &    -0.05 &   3.15 \\
tyc3854-1334-1 &          5958 &          4.32 &    -0.14 &   3.20 \\
 tyc3868-927-1 &          6436 &          4.25 &     0.11 &  19.69 \\
 tyc3946-983-1 &          6459 &          4.18 &    -0.07 &  18.51 \\
 tyc3956-693-1 &          6532 &          4.43 &    -0.15 &  16.74 \\
tyc3987-1468-1 &          6339 &          4.41 &     0.06 &  26.87 \\
 tyc4001-450-1 &          6594 &          4.34 &     0.11 &  21.66 \\
 tyc4102-698-1 &          6333 &          4.35 &    -0.24 &   2.97 \\
 tyc4105-445-1 &          6384 &          4.30 &    -0.17 &  10.45 \\
tyc4114-1545-1 &          6132 &          4.45 &    -0.21 &   3.97 \\
tyc4138-1359-1 &          6338 &          4.26 &     0.17 &   6.06 \\
 tyc4151-993-1 &          6338 &          4.33 &    -0.18 &  10.22 \\
  tyc4196-57-1 &          6126 &          4.18 &     0.00 &   3.71 \\
tyc4212-1334-1 &          6167 &          4.31 &    -0.07 &   7.08 \\
tyc4233-2300-1 &          6397 &          4.20 &     0.09 &   9.08 \\
tyc4235-1954-1 &          6291 &          4.15 &     0.26 &   7.50 \\
 tyc4298-157-1 &          6374 &          4.27 &    -0.20 &   6.49 \\
tyc4383-1286-1 &          6169 &          4.44 &    -0.16 &   5.72 \\
tyc4385-1563-1 &          5975 &          4.14 &     0.03 &   5.83 \\
 tyc4395-615-1 &          6224 &          4.23 &     0.04 &   8.62 \\
 tyc4415-128-1 &          6401 &          4.18 &    -0.17 &  24.41 \\
tyc4425-1244-1 &          6512 &          4.33 &     0.18 &  26.08 \\
 tyc4433-355-1 &          6580 &          4.33 &    -0.12 &  21.07 \\
 tyc4468-488-1 &          6464 &          4.43 &    -0.15 &   9.61 \\
 tyc4488-196-1 &          6481 &          4.40 &     0.00 &  13.44 \\
 tyc4518-605-1 &          6181 &          4.26 &     0.18 &   6.21 \\
tyc4532-1003-1 &          6490 &          4.24 &    -0.05 &  11.36 \\
tyc4541-1452-1 &          6251 &          4.11 &    -0.17 &   4.88 \\
tyc4544-1102-1 &          6321 &          4.44 &    -0.05 &  12.55 \\
  tyc4544-61-1 &          6312 &          4.39 &    -0.05 &   8.74 \\
tyc4584-1560-1 &          6334 &          4.15 &     0.03 &  14.80 \\
tyc4589-1086-1 &          6104 &          4.05 &     0.03 &   7.36 \\
 tyc4620-914-1 &          6332 &          4.36 &    -0.16 &   6.98 \\
\enddata
\end{deluxetable}

\bibliography{Keck}{}
\bibliographystyle{aasjournal}

\end{document}